# ADDING EDUCATIONAL FUNCTIONALITIES TO CLASSIC BOARD GAMES

Luis Alvarez León[1], Pablo García Tahoces[2] and Emilio Macías Conde[1]
[1]*CTIM. Departamento de Informática y Sistemas, Universidad de Las Palmas de Gran Canaria,
Campus de Tafira, 35017 Las Palmas de G.C., Spain*
[2]*CITIUS. Centro de Investigación en Tecnoloxías da Información,Universidad de Santiago de Compóstela
Campus Vida, 15782 Santiago de Compostela, Spain*
lalvarez@ctim.es, pablo.tahoces@usc.es

Abstract: In this paper we revisit some classic board games like Pachisi or the Game of Gosse. The main contribution of the paper is to design and add some functionalities to the games in order to transform them in serious games, that is, in games with learning and educational purposes. To do that, at the beginning of the game, players choose one or several topics and during the game, players have to anwers questions on these topics in order to move their markers. We choose classic board games because a lot of people are familiar with them so it is very easy to start to play without wasting time learning game rules and, we think that this is an important element to make the game more attractive to people. To enlarge the number of potential users we have implement the games just using html and javascript and the games can be used in any web browser, in any computer (including tablets) , in any computer arquitecture (Windows, Mac, Linux) and no internet/server conexion is required. Associated software is distributed under Creative Commons Attribution-NonCommercial-ShareAlike 3.0 licence and can be obtained at http://www.ctim.es/SeriousGames

## 1 INTRODUCTION

Most people are familiar with classic old board games like Pachisi or the Game of Goose. So, why do not use these games as basis to create serious games by adding some extra functionalities?. The main goal of this paper is to explore how to add educational functionalities to classic board games but keeping the game "spirit".

We have studied in details some classic board games like Game of the Goose, Parcheesi or Motorsport. All these games have the following common characteristics :

1. The game is played by several teams.
2. Each team have one of several markers that they can move in a board by throwing a dice.
3. The way a marker moves in the board follows simple rules
4. Each marker starts in a given position and follows a unique route to arrive to a given position (players can not choose where to move the marker)
5. The winner team is the one who reach first the final marker position.

Taking into account these characteristic is easy to design a software application to perform automatically marker motion according to games rules and the role of the dice. But we want to go further in this analysis. We want to add educational functionalities to the board games. To do that, at the beginning of the game, players choose one or several topics and during the game, players have to answer questions on these topics in order to move their markers. That is before moving the marker the player (or players) has to answer correctly a question randomly chosen by the system in the question database. Since the question topic database can be easily modified we can adapt the games to very different learning context. We have also included the option to add images to questions which increases a lot the learning functionalities.

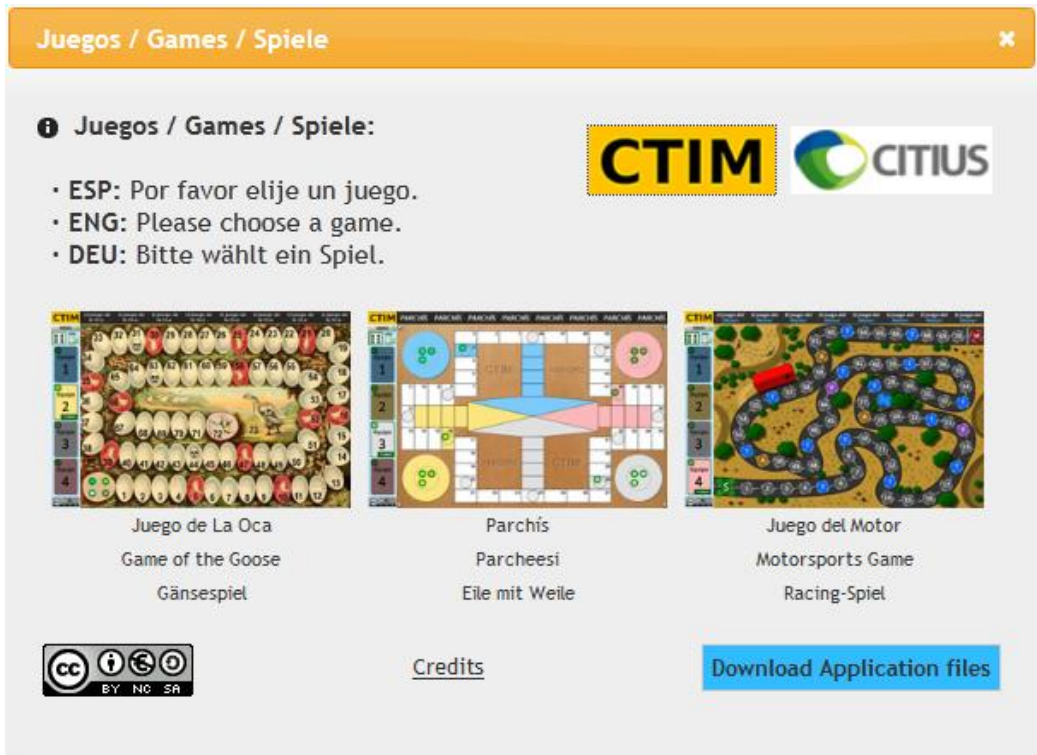

Figure 1. Main game window software application

The organization of the paper is as follow : In section 2 we study the serious game design. In section 3 we study the details of serious game software implementation and finally in section 4 we present the main conclusions of the work.

## 2  SERIOUS GAME DESIGN

The motivation we had in mind when we studied the way we could transform the classic old board games into serious games with educational purposes were the following :

1. To enjoy of classic old board games to make serious games more atractive.
2. To keep the original board game spirit in terms of basic game rules
3. The new game should be able to address a large audience by chosing the question topic level.
4. The games should run using just a web browser in order to be used by all computer arquitectures
5. Add extra functionalities in order to manage the time required to finish a match. To enlarge the number of potential users game aplication should not required internet connection to a web server. In this case, the application has to be download from http://www.ctim.es/SeriousGames (Fig 1) and its files stored in a local folder.
6. The game application is distributed under a Creative Commons licence.

### 2.1  Choice of Old Board Games

We have chosen 3 classic board games: Motorsport, Game of the Goose and Parcheesi. Next we briefly summarize the games characteristics :

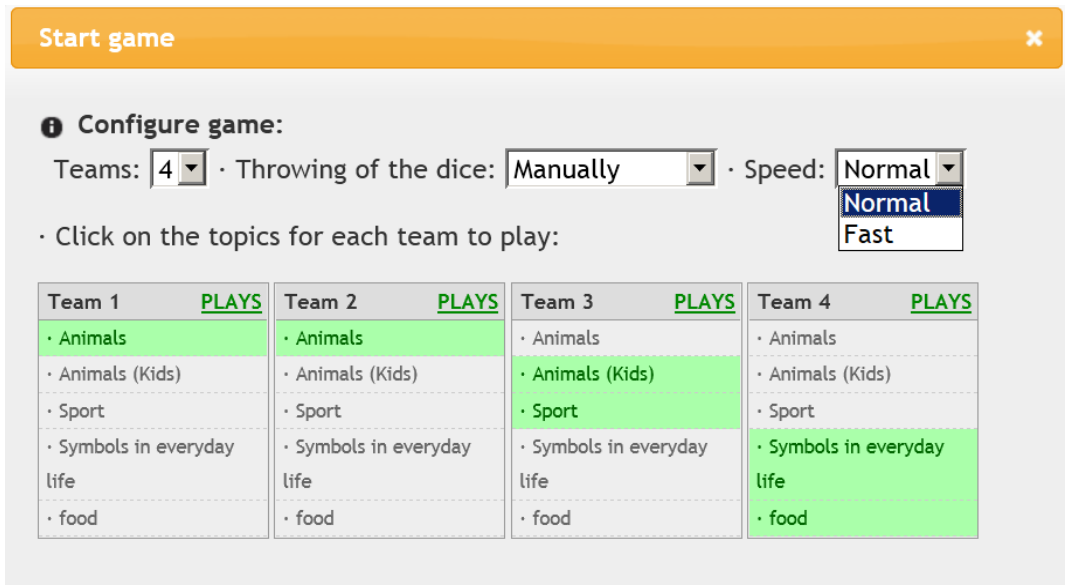

Figure 2. Game Configuration main window.

1. Motorsport : is the simple one, each team has just one marker, rules are simple and the team to reach the goal win the game. It is not necessary to reach the goal with a exact number rolling the dice.
2. Game of the Goose : technically, rules are similar to Motorsport. Teams have to reach the goal with a exact number rolling the dice.
3. Parcheesi : it is more complex than the previous ones. Teams play with 4 markers and they can decide, in any turn, the marker they move. There are a number on potential interactions between markers on the board which have to be taking into account. Teams have to reach the goal with a exact number rolling the dice.

## 2.2 Adding serious game functionalities to the Old Board Games

The way we decided to include educational functionalities to the board games is to add questions the players have to answer before moving their markers. At the beginning of the game, players choose one or several question topics. Since the topics can be different for the different teams, it is possible to play teams of children of different ages by adapting the question topic difficulties to children team age. For instance, in the English question topic database we use in this first version of the application we have included the following topics :

1. Food : identify food by their associated images
2. Animals (Kids) : identify pets or very well known animals by their associated images
3. Animals : identify animals (not always well known) by their associated images
4. Sport : Identify sports by their symbol images
5. Symbols in everyday live: Identify symbols in everyday live by their associated images

During the game, players have to answer questions on these topics in order to move their markers.

## 2.3 Managing question topics

Questions topics are managed using javascript (White, 2009) files. One very important issue is to design a procedure to help users to define their own question topics. We provide C-source code and precompiled version in windows and linux system which build automatically the javascript files from a .csv file that can be easily edited by the user (using a text editor or any spreadsheet (like excel)). The steps to add/modify the question topics are :

1. Modify the provide csv question file by adding/removing questions in the appropiate language

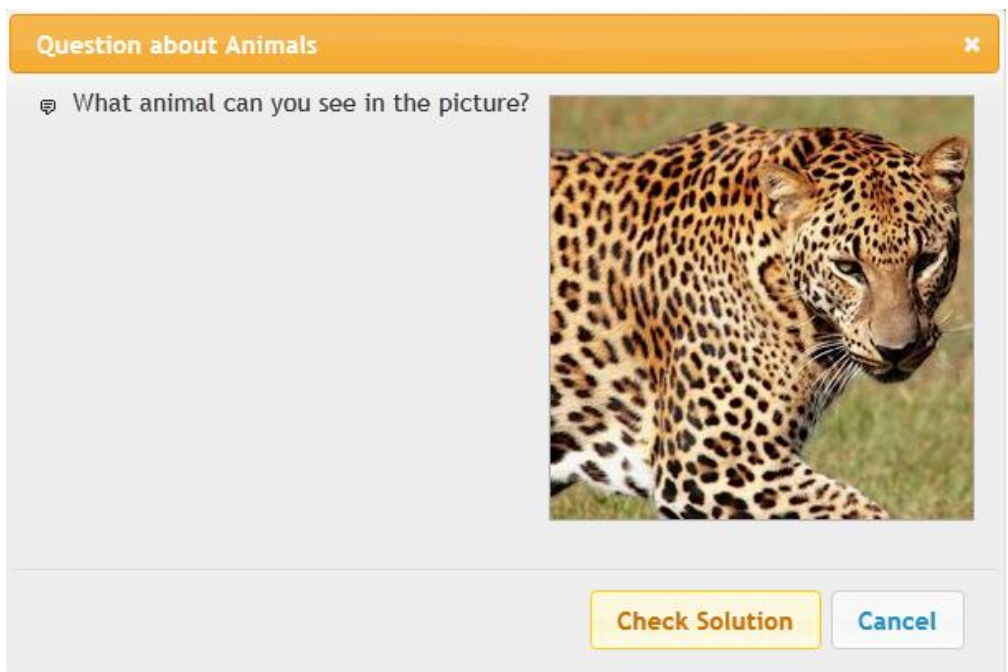

Figure 3 : Question Window Example

2. Modify Compile and execute the provided C-Source code (or use the precompiled version in Windows or Linux).
3. Replace the javascript files in the question repository by the generated new ones (taking into account the appropiate language repository).

## 2.2 Speeding up the required game time.

Quite often, especially if we want to use the application at school, the time we can devote to the game is limited, so we have to find the way to optionally speed up the game time. We have addressed this problem by adding optionally a "fast" version of the game with the following functionalities:

1. The dice numbers move randomly between 4 and 9 instead of the usual 1 and 6.
2. To arrive to the goal, it is not required an exact number rolling the dice.
3. In the case of Parcheesi, each team plays with 2 markers (instead of 4).

## 3 SERIOUS GAME SOFTWARE

The main goal we had in mind when we address the application implementation was to try to cover the maximun number of potential users. That is, the application should run in any operation system, in any hardware arquitecture and with the option of being downloaded to be run without internet conexion.

According to these requirements, we decided to implement the application using the HTML (W3C, 1999), javascript and the jquery (York, 2011) library.

HTML and javascript can be run in any web browser which provides a javascript interpreter. Currently, most of the browsers provide this functionality, but sometimes it's not activated. In that case the software cannot start.

jQuery is a JavaScript Library designed to simplify development of several dynamic effects on an html page. The basic idea of jQuery is to select HTML elements and perform some action on those elements. The basic syntax is:

```
$(selector).action()
```

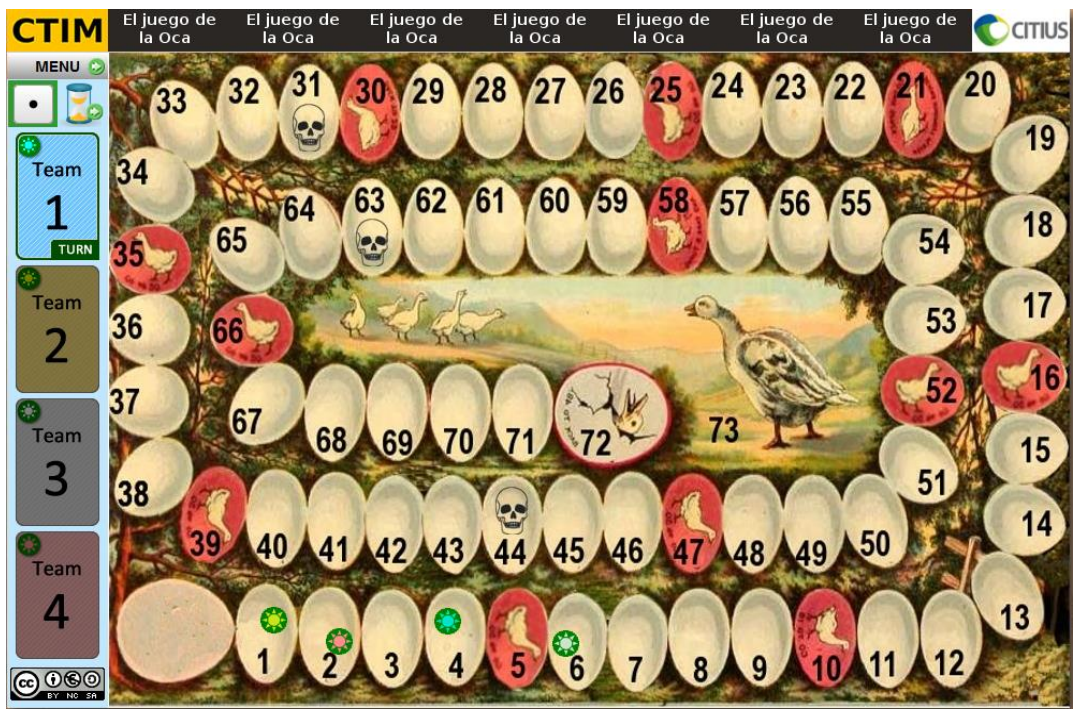

Figure 4. Game of Gosse board.

where the dollar sign ($) define a jQuery sentence, the selector allow to find HTML elements directly or by using the HTML tag "div" in a similar way as CSS (W3C, 2011) works, and the action that is performed on the elements are several methods related to events, animations, DOM (W3C, 2009) manipulation, CSS properties and AJAX (Ullman, 2007) functions.

Most of the dynamism performed on the *SeriousGame* application is created by using JQuery sentences. Following we show some examples.

i) Dialog Floating Windows

One important aspect of the game is the feasibility to interact with the player for configuration, to give information or asking for a question. All this functionality is performed via "floating windows" created with the jquery dialog() method. The dialog() method provide the possibility to generate an emerging window that include the information to show and several buttons for the user to click. The size of the window can be also controlled by the arguments of the method. Jquery dialog() have several options, but the general appearance of the method in our program is as follows:

```
$(html tag).dialog( {autopen: false,
    resizable: false,
    height: windowHeight,
      width: windowWidth,
    zIndex: zIndexNumber,
    buttons:[{//descrip. of the button
        text: "textOnTheButton",
        click:function(){javaScript
        function to run},
        { Other buttons } ]
    });
```

ii) Game Configuration

When the program starts and the language and the game has been selected, a floating window is open for selecting the number of players, the way the dice is throwing and the speed. Once those three items have been set, the program automatically decides the starting team and the game starts. To make those configurations, the html code of the pages involved has to be adapted. To perform this, jquery make use of the method html(), that set the content of a selected element:

```
$("#htmlTag").html("new html code");
```

iii) Dice Status

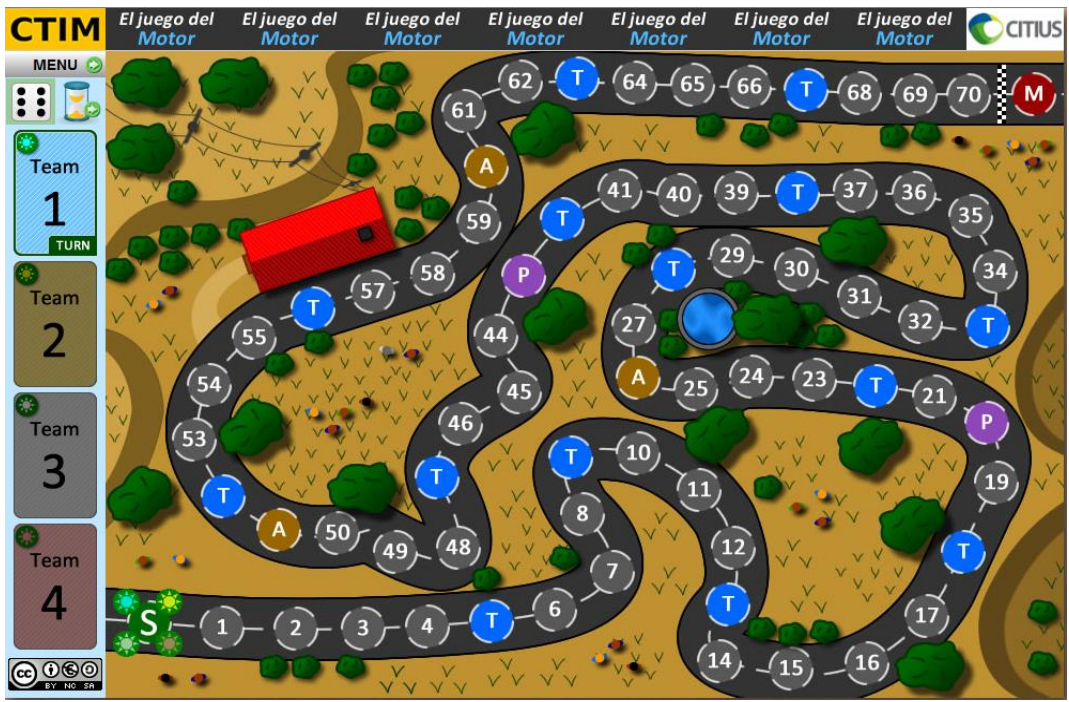

Figure 5. MotorSport game board.

The dice can have two states: locked and unlocked. Locked means that the game is waiting for the player to answer a question. In this case, the dice cannot be thrown. Unlocked means that a player owns the turn and has to thrown the dice. When the dice is not locked in the game, its background is set to green. By the contrary, when the dice is locked its background is turned red. To perform this effect we change the CSS property background-image as follows:

```
$("dice tag").css("background-image"
        ,url(locked or unlocked
        backgroundImage.jpg)");
```

iv)     Team Status

A team can have three different statuses that correspond with: team is playing, team is waiting by the turn, and team is not playing. For each status the game shows different icons. Thus, when a team is playing the icon that represents the team show the word "turn" writing in green color. When the team is not playing the message "doesn't play" is depicted in red color. Finally, when a team is waiting for a turn, no message is shown in the icon of the team. To perform this effect we change the "src" attribute from the anchor element that corresponds to the image associated to each icon as follows:

```
$("tag of the icon").attr("src",
"iconImagePath.png");
```

v)     Piece Movement

Displacement of a piece over the board game means shifting an image (the piece) from an initial position to the final position with a displacement preset speed. To represent the position two distances related with the top and the left borders of the image are employed. To perform the displacement itself the "animate()" method that include the final "position" of the object to move and the "speed" of the displacement is employed. The full sentence is writing as follows:

```
$("piece tag").animate( { left:
 posLeft, top: posTop},
displacement speed (milliseconds))
```

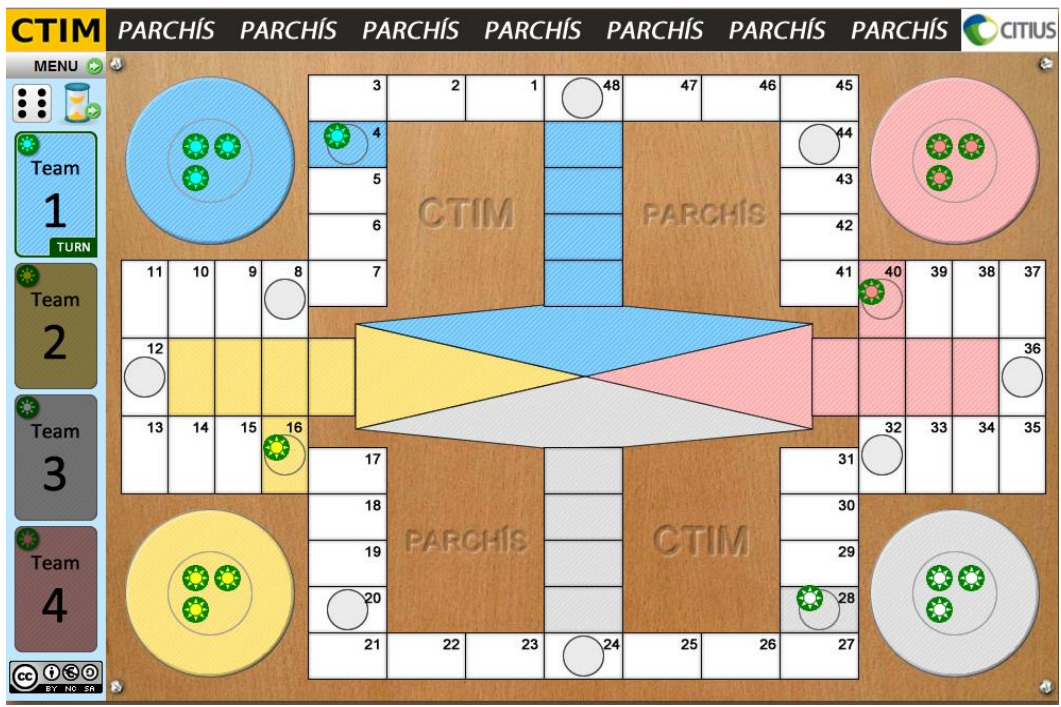

Figure 6. Pachisi game board.

## 4 CONCLUSIONS

We have designed and implemented new serious games by adding educational functionalities to classic board games.

Players can choose question topics at the beginning of the games and the application ask questions to the players before moving their markers after rolling the dice. The questions are randomly chosen from the question database.

We have used the board games : Motorsport, Game of the Goose and Parcheesi. We have implemented it using html, javascript and jquery to cover a maximun number of potential users.

The software application is distributed under the Creative Commons Attribution-NonCommercial-ShareAlike 3.0 Unported License.

Users can add easily new question topics to the application.

We think that this new approach to add educational functionalities to classic board games could be very useful at school level to learn a lot of different topics. The capability of adding images to questions also enlarge the number of potential topics we can address.